\documentclass[journal,onecolumn,11pt,draftclsnofoot]{IEEEtran}

\usepackage{cite}

\usepackage[dvips]{graphicx}
\usepackage{subfigure}
\usepackage[usenames,dvipsnames]{color}

\usepackage{amsmath,amssymb,epsfig, psfrag}
\usepackage{latexsym}

\DeclareMathOperator{\sgn}{sgn}

\makeatletter
\def\maketag@@@#1{\hbox{\m@th\normalfont\normalsize#1}} 
\makeatother

\begin{document}

\title{On the Equivalence of Maximum SNR and MMSE Estimation: Applications to \\ Additive Non-Gaussian Channels and Quantized Observations}

\author{Luca~Rugini
        and~Paolo~Banelli

\thanks{L. Rugini and P. Banelli are with the Department of Engineering, University of Perugia, 06125 Perugia, Italy (e-mail: luca.rugini@unipg.it, paolo.banelli@unipg.it)}
}

\maketitle

\begin{abstract}
The minimum mean-squared error (MMSE) is one of the most popular criteria for Bayesian estimation. Conversely, the signal-to-noise ratio (SNR) is a typical performance criterion in communications, radar, and generally detection theory. In this paper we first formalize an SNR criterion to design an estimator, and then we prove that there exists an equivalence between MMSE and maximum-SNR estimators, for any statistics. We also extend this equivalence to specific classes of suboptimal estimators, which are expressed by a basis expansion model (BEM). Then, by exploiting an orthogonal BEM for the estimator, we derive the MMSE estimator constrained to a given quantization resolution of the noisy observations, and we prove that this suboptimal MMSE estimator tends to the optimal MMSE estimator that uses an infinite resolution of the observation. Besides, we derive closed-form expressions for the mean-squared error (MSE) and for the SNR of the proposed suboptimal estimators, and we show that these expressions constitute tight, asymptotically exact, bounds for the optimal MMSE and maximum SNR.
\end{abstract}


\section{Introduction}
\IEEEPARstart{B}{ayesian} estimation of a parameter, a source, or a signal, from noisy observations, is a general framework in statistical inference, with widespread applications in signal processing, communications, controls, machine learning, etc. \cite{Kay:93}. The minimum mean-squared error (MMSE) is the most popular criterion in this framework, intuitively connected to the maximum signal-to-noise ratio (MSNR) criterion, mostly used for communication and detection applications \cite{Kay:93}, \cite{Poor:94}.
After the first seminal work in \cite{Guo:05}, the connections between the MMSE and the signal-to-noise ratio (SNR) have attracted several research interests, and there is a quite abundant literature to establish links among them and the mutual information (see \cite{Guo:08}--\cite{Yang:07} and the references therein). In the context of signal classification (i.e., detection), \cite{Gardner:80} has shown the interdependencies between the mean-squared error (MSE) and other second-order measures of quality, including many definitions of SNR. However, a thorough investigation of the links between MSE and SNR, in the context of estimation, is still lacking. Some connections between MMSE and SNR have been explored in \cite{Guo:05}, which proves that the MMSE in the additive noise channel is inversely proportional to the SNR. However, the SNR of \cite{Guo:05} is defined at the input of the estimator, while we are interested in the SNR at the output of the estimator.

Motivated to further explore the links between SNR and MSE, in this paper we first define the SNR for the output of a generic estimator, and then we prove the equivalence between the MMSE and MSNR criteria in the context of estimation design. Actually, when the parameter to be estimated and the observations are jointly Gaussian, it is well known that the MMSE estimator, the maximum likelihood (ML) estimator, and the maximum a posteriori (MAP) estimator, are linear in the observation and are equivalent to the MSNR estimator (up to a scalar multiplicative coefficient) \cite{Kay:98}, \cite{Akyol:12}: indeed, in this simple Gaussian case, all these estimators produce the same output SNR, which is both maximum and identical to the input SNR. Differently, this paper considers a more general case, where the parameter to be estimated and the observations can be non-Gaussian. In this general case, to the best of our knowledge, the natural question if the MMSE and MSNR estimation criteria are equivalent or not, is still unanswered\footnote{We believe that this question has never been addressed in detail in the context of estimation problems: the investigation done in \cite{Gardner:80} for detection cannot be extended to estimation, since the SNR definitions used in \cite{Gardner:80} are quite different from the output SNR considered in this paper.}. While classical estimation typically deals with the MMSE criterion, some authors have been looking for an MSNR solution, such as \cite{Zhidkov:06}, ignoring if this solution has anything to do with the MMSE solution. Specifically, this paper proves that the equivalence between MMSE and MSNR estimators always holds true, even when the parameter to be estimated and the observations are non-Gaussian: in this case, both the MMSE and the MSNR estimators are usually nonlinear in the observations. This equivalence establishes a strong theoretical link between MMSE and MSNR criteria, traditionally used in different contexts, i.e., estimation and detection, respectively.

Then, we prove that the equivalence between the MSNR and MMSE criteria holds true also for any suboptimal estimator that is expressed by a linear combination of fixed basis functions, according to a basis expansion model (BEM) \cite{Giannakis:98}. Within this framework, we derive the suboptimal MMSE estimator, and other equivalent MSNR estimators, constrained to a given quantization resolution of the noisy observations. Notheworthy, each quantization-constrained estimator corresponds to a specific choice of the set of BEM functions. These quantization-constrained estimators may have practical interest in low-complexity applications that use analog-to-digital (A/D) converters with limited number of bits, such as low-power wireless sensor applications. Specifically, we prove that the suboptimal quantization-constrained MMSE (Q-MMSE) estimator tends to the optimal (unquantized) MMSE estimator that uses an infinite resolution of the observation. In addition, we derive closed-form expressions for the SNR and for the MSE of the proposed suboptimal estimators. Note that these closed-form expressions can be used as lower bounds on the SNR of the MSNR estimators, or as upper bounds on the MSE of the optimal MMSE estimator: indeed, in case of non-Gaussian statistics, analytical expressions for the MMSE value are difficult to obtain \cite{Flam:12}; anyway, we also provide some analytical expressions for the MMSE and MSNR values.

To provide an example for practical applications, we apply the derived suboptimal estimators to an additive non-Gaussian noise model, where the noisy observation is simply a signal-plus-noise random variable. We include a numerical example where the signal has a Laplacian statistic, while the noise distribution is a Laplacian mixture, bearing in mind that the results in this paper are valid for any signal and noise statistics. The obtained results show that the proposed suboptimal Q-MMSE and quantization-constrained MSNR (Q-MSNR) estimators outperform other alternative estimators discussed in Section~V. The numerical results also confirm and that, when the size of the quantization intervals tends to zero, the MSE (and SNR) of the Q-MMSE estimator tends to the optimal MMSE (and MSNR) value, as expected by design. 

The rest of this paper is organized as follows. Section~II proves the equivalence between the MSNR and MMSE criteria and discusses several theoretical links. In Section~III, we derive the equivalence results for BEM-based estimators, such as the Q-MMSE. Section~IV considers the special case of additive non-Gaussian noise channel, while Section~V illustrates a numerical example. Section~VI concludes the paper.

\section{Maximum SNR and MMSE Estimators}
For real-valued scalar observation and parameters, Bayesian estimation deals with statistical inference of a random parameter of interest $x$ from a possibly noisy observation $y$, assuming that the joint probability density function (pdf) $f_{\rm{XY}}(x,y)$ is known. The estimator of the scalar parameter $x$ is a function $g(\cdot)$ that produces the estimated parameter $\hat{x} = g(y)$. By a linear regression analysis, for any zero-mean $x$ and $y$ and any estimator $g(\cdot)$, it is possible to express the estimator output as
\begin{equation}
\label{eq:lin_regr}
\hat{x} = g(y) = K_{g}x + w_g,
\end{equation}
where
\begin{equation}
\label{eq:def_k}
K_{g} =  \frac{E_{\rm{XY}}\{x g(y)\}}{\sigma^2_x},
\end{equation}
$\sigma^2_x = E_{\rm{X}}\{x^2\}$, and $w_g$ is the zero-mean output noise, which is orthogonal to the parameter of interest $x$
and characterized by $\sigma^2_{W_g} = E_{{\rm{W}}_g}\{w_g^2\}$. It is well known that the estimator $g_{\textrm{MMSE}}(\cdot)$
that minimizes the Bayesian MSE
\begin{equation}
\label{eq:def_mse}
J_{g} = E_{\rm{XY}}\{(g(y)-x)^2\}
\end{equation}
is expressed by \cite{Kay:93}, \cite{Poor:94}, \cite{Flam:12}, \cite{Banelli:13}
\begin{equation}
\label{eq:g_mmse}
g_{\textrm{MMSE}}(y) =  E_{\rm{X|Y}}\{x|y\} = \int_{-\infty}^{\infty}{x f_{\rm{X|Y}}(x|y)dx}.
\end{equation}

However, other Bayesian criteria are possible, such as the MAP, the minimum mean-absolute error, etc. \cite{Poor:94}. Actually we may choose $g(\cdot)$ that maximizes the SNR at the estimator output in \eqref{eq:lin_regr}, as done for detection in \cite{Zhidkov:06}, \cite{Zhidkov:08}. In this sense, the definition of $K_g$ in \eqref{eq:def_k} leads to the output SNR
\begin{equation}
\label{eq:def_snr}
\gamma_{g} =  \frac{K_{g}^2 \sigma^2_x}{\sigma^2_{w_g}},
\end{equation}
defined as the power ratio of the noise-free signal and the uncorrelated noise in \eqref{eq:lin_regr}. Alternatively, we may maximize the gain $K_{g}$ in \eqref{eq:def_k} (instead of the SNR), under a power constraint.

Using the orthogonality in \eqref{eq:lin_regr}, the output power is
\begin{equation}
\label{eq:output_power}
E_{\rm{Y}}\{g^2(y)\} = K^2_{g} \sigma^2_x + \sigma^2_{w_g},
\end{equation}
and hence, using \eqref{eq:def_k} and \eqref{eq:def_snr}, we obtain
\begin{align}
\label{eq:mse_vs_kg}
J_{g} =& E_{\rm{Y}}\{g^2(y)\} + (1-2K_{g})\sigma^2_x \\
\label{eq:mse_vs_kg2}
      =& (1-K_{g})^2 \sigma^2_x + \sigma^2_{w_g}.
\end{align}
From \eqref{eq:def_snr} and \eqref{eq:mse_vs_kg2}, it is straightforward that the MSE $J_{g}$ and the SNR $\gamma_g$ are linked by
\begin{equation}
\label{eq:MSEvsSNR}
J_{g} = (1-K_{g})^2 \sigma_x^2 + \frac{K_{g}^2\sigma_x^2}{\gamma_{g}}.
\end{equation}

\subsection{Equivalence of MSNR and MMSE Estimators}
While for jointly Gaussian statistics the equivalence between MSNR and MMSE is easy to establish (since the MMSE estimator is linear in $y$), herein we consider the most general case, without any assumption on the statistics of $x$ and $y$.
\vspace{3 pt}

\em Theorem 1: \em Among all the possible estimators $g(\cdot)$, the MMSE estimator \eqref{eq:g_mmse} maximizes the SNR \eqref{eq:def_snr} at the estimator output, for any pdf $f_{\rm{XY}}(x,y)$.
\vspace{3 pt}

\begin{IEEEproof}
Let us denote with $g_{\textrm{MMSE}}(y)$ the MMSE estimator \eqref{eq:g_mmse}, and with $K_{g_{\textrm{MMSE}}}$ its associated gain \eqref{eq:def_k}. In addition, let us denote with $g_{\textrm{MSNR}}(y)$ an estimator that maximizes the SNR \eqref{eq:def_snr}, as expressed by
\begin{equation}
\label{eq:x_hat_SNR_def}
g_{\textrm{MSNR}}(y)= \mathop{\arg\max}\limits_{g(\cdot)}
\left[\frac{K_{g}^2\sigma_x^2}{E_{\rm{Y}} \{g^2(y)\}-K_{g}^2\sigma_x^2}\right],
\end{equation}
and by $K_{g_{\textrm{MSNR}}}$ its associated gain in \eqref{eq:def_k}.
This MSNR estimator is not unique, since also any other estimator
\begin{equation}
g_{a,{\textrm{MSNR}}}(y) = a g_{\textrm{MSNR}}(y),
\end{equation}
with $a \in \mathbb{R} \setminus \{0\}$, maximizes the SNR.
Indeed, due to the scaling factor $a$, by means of \eqref{eq:x_hat_SNR_def} both the noise-free power $K_g^2\sigma_x^2$ and the noise power $\sigma_{w_g}^2 = E_{\rm{Y}} \{g^2(y)\}-K_{g}^2\sigma_x^2$ are multiplied by the same quantity $a^2$, hence the SNR in \eqref{eq:def_snr} is invariant with $a$. By \eqref{eq:lin_regr} and \eqref{eq:def_k}, the gain $K_{g_{a,{\textrm{MSNR}}}}$ of $g_{a,{\textrm{MSNR}}}(y)$ is equal to
\begin{equation}
\label{eq:gain_equivalence}
K_{g_{a,\textrm{MSNR}}} = a K_{g_{\textrm{MSNR}}}.
\end{equation}
Conversely, the MMSE estimator is unique and has a unique gain $K_{g_{\textrm{MMSE}}}$. Thus, we have to prove the equivalence of the MMSE estimator $g_{\textrm{MMSE}}(y)$ with the specific $g_{a,{\textrm{MSNR}}}(y)$ characterized by $K_{g_{a,\textrm{MSNR}}} = K_{g_{\textrm{MMSE}}}$. Therefore, by \eqref{eq:gain_equivalence}, we have to choose the MSNR estimator with the specific value $a=\tilde{a}$ expressed by
\begin{equation}
\label{eq:a}
\tilde{a} = \frac{K_{g_{\textrm{MMSE}}}}{K_{g_{\textrm{MSNR}}}}.
\end{equation}
The MSNR estimator $g_{\tilde{a},{\textrm{MSNR}}}(y)$ is actually the MSNR estimator that corresponds to an optimization problem restricted to the subclass of all the estimators $g(\cdot)$ characterized by the same gain $K_g=K_{g_{\textrm{MMSE}}}$, as expressed by
\begin{equation}
\label{eq:g_MSNR_kmmse}
g_{\tilde{a},\textrm{MSNR}}(y)= \mathop{\arg\max}\limits_{g(\cdot), K_g = K_{g_{\textrm{MMSE}}}} 
\left[\frac{{K^2_g}\sigma_x^2}{E_{\rm{Y}} \{g^2(y)\}-{K^2_g}\sigma_x^2}\right].
\end{equation}
Note that, despite the constraint $K_g = K_{g_{\textrm{MMSE}}}$, we still obtain the unconstrained MMSE estimator \eqref{eq:g_mmse}, which by definition belongs to the subclass of estimators being characterized by $K_g=K_{g_{\textrm{MMSE}}}$. Using the constraint $K_g = K_{g_{\textrm{MMSE}}}$, it is clear in \eqref{eq:MSEvsSNR} that the dependence of the MSE functional $J_g$ on $g(\cdot)$ is only through $\gamma_g $, and no longer also through $K_g$ as in the general case: consequently, the MMSE estimator is
\begin{align}
g_{\textrm{MMSE}}(y)= & \mathop{\arg \min}\limits_{g(\cdot), K_g = K_{g_{\textrm{MMSE}}}}
\left[J_{g}\right] = \mathop {\arg \min }\limits_{g(\cdot), K_g = K_{g_{\textrm{MMSE}}}}\left[
\frac{\sigma_x^2}{\gamma_g}\right] \nonumber \\
=& \mathop {\arg \max }\limits_{g(\cdot), K_g = K_{g_{\textrm{MMSE}}}}\left[
{\gamma_g}\right] = g_{\tilde{a},\textrm{MSNR}}(y).
\label{eq:g_MMSE_kmmse}
\end{align}
Thus, \eqref{eq:g_MMSE_kmmse} shows that the estimator that maximizes the SNR with a fixed $K_g=K_{g_{\textrm{MMSE}}}$ is equivalent to the estimator that minimizes the MSE, i.e.,  $g_{\tilde{a},\textrm{MSNR}}(y)$ = $g_{\textrm{MMSE}}(y)$.
\end{IEEEproof}
\vspace{3 pt}

Basically, Theorem 1 explains that $\{g_{a,{\textrm{MSNR}}}(y)\}$ are all scaled versions of $g_{\textrm{MMSE}}(y)$. In other words, each scaled version of the MSNR produces the same SNR, but a different MSE: only a unique MSNR estimator is the MMSE estimator, and, in this sense, the two estimation criteria are equivalent.

\subsection{Theoretical Properties of  MSNR and MMSE Estimators}

\em Property 1: \em The output power $E_{\rm{Y}} \{g_{\textrm{MMSE}}^2(y)\}$ of the MMSE estimator \eqref{eq:g_mmse} is equal to $K_{g_{\textrm{MMSE}}} \sigma_x^2$. Indeed, from \eqref{eq:def_k} and \eqref{eq:g_mmse}, we obtain
\begin{align}
K_{g_{\textrm{MMSE}}} \sigma_x^2 =&  E_{\rm{XY}}\{x g_{\textrm{MMSE}}(y)\} =  E_{\rm{XY}}\{x E_{\rm{X|Y}}\{x|y\} \}  \nonumber \\
                                 =&  \int_{-\infty}^{\infty}{ E_{\rm{X|Y}}\{x E_{\rm{X|Y}}\{x|y\} | y \} f_{\rm{Y}}(y) dy} \nonumber \\
                                 =&  \int_{-\infty}^{\infty}{ \left[ E_{\rm{X|Y}}\{x|y\} \right] ^2 f_{\rm{Y}}(y) dy} \nonumber \\
                                 =& E_{\rm{Y}} \{g_{\textrm{MMSE}}^2(y)\}.
\label{eq:output_power_mmse}
\end{align}

\em Property 2: \em The MMSE $J_{g_{\textrm{MMSE}}}$ is equal to $(1 - K_{g_{\textrm{MMSE}}}) \sigma_x^2$. Indeed, from \eqref{eq:mse_vs_kg} and \eqref{eq:output_power_mmse}, we obtain
\begin{align}
J_{g_{\textrm{MMSE}}} =& E_{\rm{Y}}\{g_{\textrm{MMSE}}^2(y)\} + (1-2K_{g_{\textrm{MMSE}}})\sigma^2_x \nonumber \\
                      =& (1-K_{g_{\textrm{MMSE}}})\sigma^2_x. 
\label{eq:mmse_value}
\end{align}

\em Property 3: \em The power of the uncorrelated noise term $w_g$ at the output of the MMSE estimator is equal to $ K_{g_{\textrm{MMSE}}}(1 - K_{g_{\textrm{MMSE}}})\sigma^2_x $. Indeed, from \eqref{eq:output_power}, \eqref{eq:output_power_mmse}, and \eqref{eq:mmse_value}, we obtain
\begin{align}
\sigma^2_{w_{g_{\textrm{MMSE}}}} =& E_{\rm{Y}}\{g_{\textrm{MMSE}}^2(y)\} - K_{g_{\textrm{MMSE}}}^2 \sigma^2_x \nonumber \\
\label{eq:noise_power_mmse}
           =& K_{g_{\textrm{MMSE}}} ( 1 - K_{g_{\textrm{MMSE}}}) \sigma_x^2 \\ 
           =& K_{g_{\textrm{MMSE}}} J_{g_{\textrm{MMSE}}}. \nonumber 
\end{align}
Equation \eqref{eq:noise_power_mmse} confirms that $K_{g_{\textrm{MMSE}}} \in [0,1]$.

\em Property 4: \em The MSNR $\gamma_{g_{\textrm{MSNR}}}$ is equal to $ K_{g_{\textrm{MMSE}}} / (1 - K_{g_{\textrm{MMSE}}})$. Indeed, from \eqref{eq:def_snr} and \eqref{eq:noise_power_mmse}, we obtain
\begin{equation}
\label{eq:msnr_value}
\gamma_{g_{\textrm{MSNR}}} = \gamma_{g_{\textrm{MMSE}}} = \frac{K_{g_{\textrm{MMSE}}}^2 \sigma_x^2}{\sigma^2_{w_{g_{\textrm{MMSE}}}}} = \frac{K_{g_{\textrm{MMSE}}}}{1 - K_{g_{\textrm{MMSE}}}}.
\end{equation}
By \eqref{eq:output_power_mmse}--\eqref{eq:msnr_value}, the MSNR is related to the MMSE by
\begin{equation}
\label{eq:msnr_vs_mmse}
\gamma_{g_{\textrm{MSNR}}} = \gamma_{g_{\textrm{MMSE}}} = \frac{E_{\rm{Y}} \{g_{\textrm{MMSE}}^2(y)\}}{J_{g_{\textrm{MMSE}}}} = \frac{\sigma^2_x - J_{g_{\textrm{MMSE}}}}{J_{g_{\textrm{MMSE}}}}.
\end{equation}

\em Property 5: \em The unbiased MMSE (UMMSE) estimator $g_{\textrm{UMMSE}}(y)$ maximizes the SNR: therefore, the UMMSE estimator is a scaled version of the MMSE estimator, i.e.,
\begin{equation}
\label{eq:g_ummse}
g_{\textrm{UMMSE}}(y) = \frac{g_{\textrm{MMSE}}(y)}{K_{g_{\textrm{MMSE}}}}.
\end{equation}
Indeed, for any estimator $g(y)$, we can make it unbiased by dividing $g(y)$ by $K_g$, as expressed by
\begin{equation}
\label{eq:lin_regr_unbiased}
\hat{x} = h(y) = \frac{g(y)}{K_g} = x + \frac{w_g}{K_g}.
\end{equation}
By \eqref{eq:lin_regr}, $ h(y) = K_h x + w_h$, therefore $K_h = 1$ and $w_h = {w_g}/{K_g}$. Hence, for unbiased estimators, the minimization over $h(\cdot)$ of the MSE $\sigma_{w_h}^2$ is equivalent to the minimization over $g(\cdot)$ of $\sigma_{w_g}^2 / {K_g^2}$, which coincides with the maximization over $g(\cdot)$ of the SNR \eqref{eq:def_snr}. As a consequence, the UMMSE estimator is the unique MSNR estimator characterized by $K_{g_{\textrm{MSNR}}} = 1$. Since all MSNR estimators are scaled versions of $g_{\textrm{MMSE}}(y)$, the unique UMMSE estimator coincides with \eqref{eq:g_ummse}.

\em Property 6: \em The MSE $J_{g_{\textrm{UMMSE}}}$ of the UMMSE estimator is equal to $J_{g_{\textrm{MMSE}}} /K_{g_{\textrm{MMSE}}}$. Indeed, from \eqref{eq:g_ummse}, \eqref{eq:output_power_mmse}, and \eqref{eq:mmse_value}, it is easy to show that
\begin{equation}
\label{eq:mse_value_unbiased}
J_{g_{\textrm{UMMSE}}} = \frac{\sigma_{w_{g_{\textrm{MMSE}}}}^2}{K_{g_{\textrm{MMSE}}}^2} = \frac{1-K_{g_{\textrm{MMSE}}}}{K_{g_{\textrm{MMSE}}}}\sigma_x^2 = \frac{J_{g_{\textrm{MMSE}}}}{K_{g_{\textrm{MMSE}}}}.
\end{equation}
Since $K_{g_{\textrm{MMSE}}} \le 1$, then $J_{g_{\textrm{UMMSE}}} \ge J_{g_{\textrm{MMSE}}}$.
\vspace{3 pt}

The Properties 1-6, summarized in Table~\ref{table_properties}, show that all the theoretical expressions for both MMSE and MSNR basically depend on $K_{g_{\textrm{MMSE}}}$. Since the definition of $K_{g_{\textrm{MMSE}}}$ in \eqref{eq:def_k} involves a double integration over the joint pdf $f_{\rm{XY}}(x,y)$, in general the exact value of $K_{g_{\textrm{MMSE}}}$ is difficult to obtain analytically. Hence, we introduce some suboptimal estimators that allow for an analytical evaluation of their MSE and SNR.

\begin{table}[t]
\centering
\caption{Summary of Theoretical Properties}
\label{table_properties}
  \def\arraystretch{1.5}
  \begin{tabular}{  c | c | c  }
    \hline
    \# & Meaning & Expression \\
    \hline
    \hline
    1 & MMSE output power  & $E_{\rm{Y}} \{g_{\textrm{MMSE}}^2(y)\} = K_{g_{\textrm{MMSE}}} \sigma_x^2$ \\
    \hline
    2 & MMSE value         & $J_{g_{\textrm{MMSE}}} = (1 - K_{g_{\textrm{MMSE}}}) \sigma_x^2$ \\
    \hline
    3 & MMSE output noise  & $\sigma^2_{w_{g_{\textrm{MMSE}}}} = K_{g_{\textrm{MMSE}}}(1 - K_{g_{\textrm{MMSE}}})\sigma^2_x$ \\
    \hline
    4 & MSNR value         & $\gamma_{g_{\textrm{MSNR}}} = K_{g_{\textrm{MMSE}}} / (1 - K_{g_{\textrm{MMSE}}})$ \\
    \hline
    5 & UMMSE estimator    & $g_{\textrm{UMMSE}}(y) = {g_{\textrm{MMSE}}(y)}/{K_{g_{\textrm{MMSE}}}}$ \\
    \hline
    6 & MSE of UMMSE estim.& $J_{g_{\textrm{UMMSE}}} = J_{g_{\textrm{MMSE}}} /K_{g_{\textrm{MMSE}}}$ \\
    \hline
  \end{tabular}
\end{table}

\section{Suboptimal Estimators}
Suboptimal MMSE and MSNR estimators for non-Gaussian statistics are interesting for several reasons. For instance, closed-form computation of the MMSE estimator $g_{\textrm{MMSE}}(y)$ in \eqref{eq:g_mmse} may be cumbersome. Furthermore, the optimal MMSE nonlinear function $g_{\textrm{MMSE}}(y)$ may be too complicated to be implemented by low-cost hardware, such as wireless sensors. Additionally, the MMSE $J_{g_{\textrm{MMSE}}}$ is difficult to compute in closed form. Consequently, a simpler analytical expression for a suboptimal estimator $g(\cdot)$ may permit to compute the associated MSE and SNR, which provide an upper bound on the MMSE and a lower bound on the MSNR, respectively.

Considering a wide class of suboptimal estimators, we assume that $g(\cdot)$ is expressed by a BEM of $N$ known functions $u_i(\cdot)$ and $N$ unknown coefficients $g_i$:
\begin{equation}
\label{eq:g_bem}
g(y) =  \sum_{i=1}^{N}{g_i u_i(y)}.
\end{equation}
Each function $u_i(y)$ can be interpreted as a specific (possibly highly suboptimal) estimator, and $g(y)$ in \eqref{eq:g_bem} as a linear combination of simpler estimators. We are not interested in the optimization of the basis functions $\{u_i(\cdot)\}$: therefore, the design of $g(\cdot)$ becomes the design of the coefficients $\{g_i\}$. Actually, we have no constraints on the choice of $\{u_i(\cdot)\}$; for instance, saturating or blanking functions, or a mix of them, are typically beneficial to contrast impulsive noise \cite{Zhidkov:06}, \cite{Zhidkov:08}. However, in Section~III.C, we will show that an orthogonal design simplifies the computation of $\{g_i\}$, and that the proposed design is general enough for any context. 

In the following two subsections, we show that, for BEM-constrained suboptimal estimators \eqref{eq:g_bem}, the MSNR and MMSE design criteria still continue to be equivalent.

\subsection{B-MSNR Estimators}
Herein we derive the MSNR estimators constrained to the BEM \eqref{eq:g_bem}, denoted as BEM-MSNR (B-MSNR) estimators. By \eqref{eq:output_power} and \eqref{eq:g_bem}, the SNR $\gamma_{g}$ in \eqref{eq:def_snr} can be expressed by
\begin{equation}
\label{eq:snr_1}
\gamma_{g} = \frac{{K_{g}^2} \sigma^2_x}{E_{\rm{Y}}\{g^2(y)\} - {K_{g}^2} \sigma^2_x} =
\frac{\mathbf{g}^T \boldsymbol{\theta} \boldsymbol{\theta}^T
\mathbf{g}}{\mathbf{g}^T (\sigma^2_x \mathbf{R} - \boldsymbol{\theta}
\boldsymbol{\theta}^T) \mathbf{g}}.
\end{equation}
where
\begin{align}
\label{eq:def_g}
\mathbf{g} =& [g_1, g_2, ..., g_N]^T, \\
\label{eq:def_theta}
\boldsymbol{\theta} =& [\theta_1, \theta_2, ..., \theta_N]^T, \\
\label{eq:def_theta_i}
\theta_i =& E_{\rm{XY}}\{x u_i(y)\}, \\
\label{eq:def_phi}
\mathbf{R} =& \begin{bmatrix} R_{11} & \cdots & R_{1N} \\ \vdots &
\ddots & \vdots \\ R_{N1} & \cdots & R_{NN} \end{bmatrix}, \\
\label{eq:def_phi_ij}
R_{ij} =& E_{\rm{Y}}\{u_i(y) u_j(y)\}.
\end{align}

In order to maximize \eqref{eq:snr_1}, we take the eigenvalue decomposition of the symmetric matrix $\sigma^2_x \mathbf{R} - \boldsymbol{\theta} \boldsymbol{\theta}^T = \mathbf{U} \boldsymbol{\Lambda} \mathbf{U}^T$, which is assumed to be full rank. Note that $\mathbf{U}$ is orthogonal and $\boldsymbol{\Lambda}$ is diagonal. Then, we express the SNR in \eqref{eq:snr_1} as
\begin{equation}
\label{eq:snr_2}
\gamma_{g} = \frac{\mathbf{v}^T \mathbf{b} \mathbf{b}^T \mathbf{v}}{\mathbf{v}^T \mathbf{v}},
\end{equation}
where $\mathbf{v} = \boldsymbol{\Lambda}^{1/2} \mathbf{U}^T \mathbf{g}$ and $\mathbf{b} = \boldsymbol{\Lambda}^{-1/2} \mathbf{U}^T \boldsymbol{\theta}$. The ratio in \eqref{eq:snr_2} is maximum \cite{Golub:13} when $\mathbf{v} = \mathbf{v}_{\textrm{B-MSNR}} = c \mathbf{b} = c \boldsymbol{\Lambda}^{-1/2} \mathbf{U}^T \boldsymbol{\theta}$, where $c \in \mathbb{R} \setminus \{0\}$ is an arbitrary constant, and therefore the SNR in \eqref{eq:snr_1} is maximum when the estimator is
\begin{equation}
\label{eq:g_msnr}
\mathbf{g}_{\textrm{B-MSNR}} = \mathbf{U} \boldsymbol{\Lambda}^{-1/2} \mathbf{v}_{\textrm{B-MSNR}} = c (\sigma^2_x \mathbf{R} - \boldsymbol{\theta} \boldsymbol{\theta}^T)^{-1} \boldsymbol{\theta}.
\end{equation}

By \eqref{eq:snr_1} and \eqref{eq:g_msnr}, using the Sherman-Morrison formula \cite{Golub:13}, the SNR of B-MSNR estimators is expressed by
\begin{equation}
\label{eq:snr_msnr}
\gamma_{\textrm{B-MSNR}} = \boldsymbol{\theta}^T (\sigma^2_x \mathbf{R} -
\boldsymbol{\theta} \boldsymbol{\theta}^T)^{-1} \boldsymbol{\theta} =
\frac{\boldsymbol{\theta}^T \mathbf{R}^{-1} \boldsymbol{\theta}}{\sigma^2_x
-
\boldsymbol{\theta}^T \mathbf{R}^{-1} \boldsymbol{\theta}}.
\end{equation}

\subsection{B-MMSE Estimator}
Now we derive the MMSE estimator constrained to the BEM \eqref{eq:g_bem}, denoted as BEM-MMSE (B-MMSE) estimator. By \eqref{eq:g_bem} and \eqref{eq:def_g}--\eqref{eq:def_phi_ij}, the MSE
$J_{g}$ in \eqref{eq:mse_vs_kg2} becomes
\begin{equation}
\label{eq:mse_2}
J_g = \sigma^2_x -2 \mathbf{g}^T \boldsymbol{\theta} + \mathbf{g}^T \mathbf{R}
\mathbf{g}.
\end{equation}
By taking the derivative of \eqref{eq:mse_2} with respect to $\mathbf{g}$ and setting it to zero, we obtain
the B-MMSE estimator, expressed by
\begin{equation}
\label{eq:g_mmse_constr}
\mathbf{g}_{\textrm{B-MMSE}} = \mathbf{R}^{-1} \boldsymbol{\theta}.
\end{equation}
By \eqref{eq:mse_2} and \eqref{eq:g_mmse_constr}, the MSE of the B-MMSE estimator is
\begin{equation}
\label{eq:mse_mmse}
J_{g_{\textrm{B-MMSE}}} = \sigma^2_x - \mathbf{g}_{\textrm{B-MMSE}}^T \mathbf{R} \mathbf{g}_{\textrm{B-MMSE}} = \sigma^2_x - \boldsymbol{\theta}^T \mathbf{R}^{-1} 
\boldsymbol{\theta}.
\end{equation}
Using \eqref{eq:mse_mmse}, the SNR \eqref{eq:snr_msnr} can be expressed by
\begin{equation}
\label{eq:mse_snr_bem}
\gamma_{\textrm{B-MSNR}} = \frac{\boldsymbol{\theta}^T \mathbf{R}^{-1} \boldsymbol{\theta}}{J_{g_{\textrm{B-MMSE}}}} = \frac{\sigma^2_x - J_{g_{\textrm{B-MMSE}}}}{J_{g_{\textrm{B-MMSE}}}}.
\end{equation}
The similarity of \eqref{eq:mse_snr_bem} and \eqref{eq:msnr_vs_mmse} suggests a link between B-MMSE and B-MSNR estimators, as shown in Theorem 2.
\vspace{3 pt}

\em Theorem 2: \em The B-MSNR estimator \eqref{eq:g_msnr}
coincides with the B-MMSE estimator \eqref{eq:g_mmse_constr}, when $c = \sigma^2_x - \boldsymbol{\theta}^T \mathbf{R}^{-1} \boldsymbol{\theta}$.

\begin{IEEEproof}
Using the Sherman-Morrison formula \cite{Golub:13}, \eqref{eq:g_msnr} becomes
\begin{equation}
\label{eq:g_msnr_1}
\mathbf{g}_{\textrm{B-MSNR}} = \frac{c }{\sigma^2_x - \boldsymbol{\theta}^T
\mathbf{R}^{-1} \boldsymbol{\theta}} \mathbf{R}^{-1} \boldsymbol{\theta}.
\end{equation}
When $c = \sigma^2_x - \boldsymbol{\theta}^T \mathbf{R}^{-1}
\boldsymbol{\theta}$, $\mathbf{g}_{\textrm{B-MSNR}}$ in \eqref{eq:g_msnr_1} coincides
with $\mathbf{g}_{\textrm{B-MMSE}}$ in \eqref{eq:g_mmse_constr}.
\end{IEEEproof}
\vspace{3 pt}

Theorem 2 proves that the B-MMSE estimator maximizes the SNR \eqref{eq:snr_1} among all the BEM-based estimators: therefore, each B-MSNR estimator is a scaled version of the B-MMSE estimator. Also in this BEM-constrained case the equivalence between B-MMSE and B-MSNR estimators is valid for any statistic of the signal and of the noisy observation.

Note that in Theorem 2 the functions $\{u_i(\cdot)\}$ are arbitrary, but fixed. Differently, if we fix the coefficients $\{g_i\}$ in \eqref{eq:g_bem}, and perform the optimization over a subset of functions, the equivalence between MMSE and MSNR solutions may not hold true. Indeed, in case of impulsive noise mitigation by means of a soft limiter (SL), expressed by $g_{\textrm{SL}}(y) = -\beta$ if $y \leq -\beta$, $g_{\textrm{SL}}(y) = y$ if $-\beta < y < \beta$, and $g_{\textrm{SL}}(y) = \beta$ if $y \geq \beta$, the optimization over $\beta > 0$ generally produces an MMSE solution \cite{Banelli:13} that is different from the MSNR solution \cite{Zhidkov:08}. Therefore, the equivalence between MMSE and MSNR estimators can be invalid for non-BEM-based suboptimal estimators.

In addition to MMSE, there exist other criteria that maximize the SNR: as shown in Appendix~A, the BEM-based unbiased MMSE estimator and a BEM-based estimator that maximizes the gain \eqref{eq:def_k} (subject to a power constraint) both produce the same SNR of B-MMSE and B-MSNR estimators.

\subsection{Q-MMSE Estimator}
Herein we prove that, by choosing convenient basis functions $\{ u_i(\cdot) \}$ in \eqref{eq:g_bem}, the B-MMSE estimator \eqref{eq:g_mmse_constr} converges to the optimal MMSE estimator \eqref{eq:g_mmse}. Indeed, the rectangular disjoint (orthogonal) basis functions
\begin{equation}
\label{eq:rect}
u_i(y) = \begin{cases} 1, & \mbox{if } y_{i-1} < y \leq y_i,\\ 0, & \mbox{otherwise,} \end{cases}
\end{equation}
for $i=1,...,N$, with $y_0 = -\infty$ and $y_N = \infty$, greatly simplify the computation of the coefficients $\{g_i\}$. Basically, we are approximating the estimator $g(y)$ by a piecewise-constant function. Using \eqref{eq:rect}, $R_{ij}$ in \eqref{eq:def_phi_ij} becomes
\begin{equation}
\label{eq:def_phi_ij_1}
R_{ij} = \begin{cases}F_{\rm{Y}}(y_i) - F_{\rm{Y}}(y_{i-1}), & \mbox{if } i=j,\\ 0, & \mbox{if } i \ne j, \end{cases}
\end{equation}
where $F_{\rm{Y}}(y)$ is the cumulative distribution function (cdf) of the observation $y$. In this case, the matrix $\mathbf{R}$ in \eqref{eq:def_phi} is diagonal. Therefore, the coefficients of this specific B-MMSE estimator \eqref{eq:g_mmse_constr}, which we refer to as Q-MMSE estimator, simply become
\begin{equation}
\label{eq:g_i_mmse}
g_{i,\textrm{Q-MMSE}} = \frac{\theta_i}{R_{ii}},
\end{equation}
while the associated MSE \eqref{eq:mse_mmse} is expressed by
\begin{equation}
\label{eq:mse_mmse_2}
J_{g_{\textrm{Q-MMSE}}} = \sigma^2_x - \sum_{i=1}^{N} {\frac{\theta_i^2}{R_{ii}}}.
\end{equation}

Note that the Q-MMSE estimator \eqref{eq:g_i_mmse} can also be interpreted as the MMSE estimator when the observation $y$ has been discretized using $N$ quantization intervals $( y_{i-1} , y_{i} ]$, for $i = 1,...,N$. Moreover, we should bear in mind that the number $N$ of quantization levels, as well as the edges of the quantization intervals, are fixed but arbitrary. Thus, the proposed framework finds a natural application when the observed signal undergoes an A/D conversion stage.

However, it is important to prove that, in case of infinite number of quantization levels, the Q-MMSE estimator \eqref{eq:g_i_mmse} tends to the optimal MMSE estimator \eqref{eq:g_mmse} for unquantized observations: hence, the number $N$ of quantization levels enables a tradeoff between performance and complexity.
\vspace{3 pt}

\em Theorem 3: \em When the interval size $\Delta y_i = y_i - y_{i-1}$ tends to
zero for $i=2,...,N-1$, and when $y_1$ and $y_{N-1}$ tend to $y_0 = - \infty$
and $y_N = \infty$, respectively, then the Q-MMSE estimator \eqref{eq:g_i_mmse}
tends to the MMSE estimator \eqref{eq:g_mmse}.
\vspace{3 pt}

\begin{IEEEproof}
When $\Delta y_i \rightarrow 0$, for $i=2,...,N-1$, from \eqref{eq:rect} it is
easy to show that $f_{\rm{X|Y}}(x|y) u_i(y) \rightarrow f_{\rm{X|Y}}(x|y_i) u_i(y)$;
hence, for $i=2,...,N-1$, \eqref{eq:def_theta_i} gives
\begin{align}
\label{eq:theta_delta}
\theta_i =&  \int_{-\infty}^{\infty}{x \int_{y_{i-1}}^{y_i} {f_{\rm{X|Y}}(x|y)
f_{\rm{Y}}(y) dy} dx} \xrightarrow{\Delta y_i \to 0} \nonumber \\
\rightarrow & f_{\rm{Y}}(y_i) \Delta y_i \int_{-\infty}^{\infty}{x
f_{\rm{X|Y}}(x|y_i)dx}.
\end{align}
In addition, from \eqref{eq:rect} and \eqref{eq:def_phi_ij_1}, we have
\begin{equation}
\label{eq:phi_delta}
R_{ii} =  \int_{y_{i-1}}^{y_i} {f_{\rm{Y}}(y) dy} \xrightarrow{\Delta y_i \to 0}
f_{\rm{Y}}(y_i)
\Delta y_i.
\end{equation}
By taking the ratio between \eqref{eq:theta_delta} and \eqref{eq:phi_delta},
$g_{i,\textrm{Q-MMSE}}$ in \eqref{eq:g_i_mmse} tends to $g_{\textrm{MMSE}}(y_i)
= E_{\rm{X|Y}}\{x|y_i\}$ in \eqref{eq:g_mmse}, for $i=2,...,N-1$. This
result can be extended in order to include $i=1$ and $i=N$ by noting that,
when $y_1 \rightarrow y_0 = - \infty$ and $y_{N-1} \rightarrow y_N = \infty$,
then $f_{\rm{X|Y}}(x|y) \rightarrow f_{\rm{X|Y}}(x|y_1)$ for $y \in (y_0,y_1]$
and $f_{\rm{X|Y}}(x|y) \rightarrow f_{\rm{X|Y}}(x|y_{N-1})$ for $y \in (y_{N-1},y_N)$.
\end{IEEEproof}
\vspace{3 pt}

Theorem 3 proves that, when the size of the quantization intervals tends to zero, the Q-MMSE estimator converges to the MMSE estimator, regardless of the statistics of the signal of interest $x$ and of the noisy observation $y$. In particular, the SNR of the Q-MMSE estimator converges to the SNR of the MMSE estimator. Moreover, since a Q-MMSE estimator is a particular B-MMSE estimator, by Theorem 2, the Q-MMSE estimator is also a Q-MSNR estimator, for the same set of quantization thresholds. Noteworthy, if we would optimize the quantization intervals $\{ (y_{i-1},y_i] \}$ [i.e., the functions $\{u_i(\cdot)\}$ in \eqref{eq:rect}] by keeping the coefficients $g_i$ as fixed, we could end up with different quantization thresholds in an MMSE and MSNR sense.

\section{Q-MMSE in Additive Noise Channels}

\begin{figure} [t]
\centerline{\includegraphics[width=0.7\textwidth]{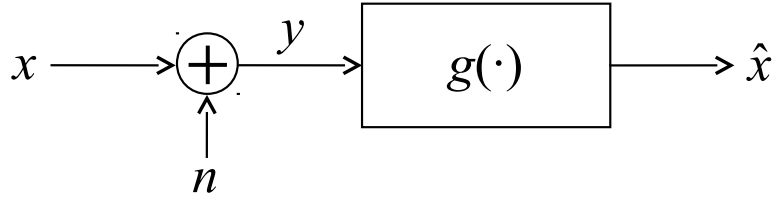}}
\caption{Signal estimation in additive noise channels.}
\label{fig_syst_mod}
\end{figure}

Herein we provide further insights on the coefficients \eqref{eq:g_i_mmse} of the Q-MMSE estimator, when the observations are impaired by an additive noise $n$, independent from $x$, as expressed by
\begin{equation}
\label{eq:noise_model}
y = x + n
\end{equation}
and depicted in Fig.~\ref{fig_syst_mod}. The additive noise model \eqref{eq:noise_model} occurs in several applications, especially if the data are obtained by quantized measurements. Indeed, Q-MMSE estimators are particularly useful in realistic scenarios where either the source, or the noise, or both, depart from the standard Gaussian assumption. These scenarios include: (a) additive noise with a high level of impulsiveness \cite{Friedmann:00}--\cite{Juwono:15}; (b) additive noise whose pdf is a mixture of statistics caused by the random occurrence of different noise sources \cite{Chuah:00}--\cite{Beaulieu:10}; (c) source represented by a pdf mixture, such as in applications (e.g., audio, medical, etc.) that involve effective denoising of sounds or images \cite{Rabbani:09, Rabbani:10}. The optimal coefficients $\{g_i\}$ obviously depend on the specific pdfs of source and noise, and the numerical results reported in Section~V give some evidence of the usefulness of Q-MMSE estimation in an additive non-Gaussian observation model.

According to the BEM model, we assume that the quantization thresholds have been fixed by some criterion. Despite possible criteria for threshold optimization are beyond the scope of this work, in Section~V we give some insights about this issue and consider some heuristic solutions.

To specialize the results of Section~III to the additive noise model in \eqref{eq:noise_model}, we observe that the pdf $f_{\rm{Y}}(y)$ is the convolution between $f_{\rm{X}}(x)$ and $f_{\rm{N}}(n)$. Thus, the coefficients $\theta_i$ and $R_{ii}$ defined in \eqref{eq:def_theta_i} and \eqref{eq:def_phi_ij} can be calculated from the first-order statistics of $x$ and $n$. Using \eqref{eq:noise_model}, \eqref{eq:def_theta_i} and \eqref{eq:rect}, we obtain
\begin{equation}
\label{eq:theta_i_noise_1}
\theta_i = \int_{-\infty}^{\infty}{x f_{\rm{X}}(x) \int_{y_{i-1}-x}^{y_i-x}
{f_{\rm{N}}(n) dn} dx} = D(y_i) - D(y_{i-1}),
\end{equation}
where
\begin{equation}
\label{eq:d_1}
D(y) = \int_{-\infty}^{\infty}{x f_{\rm{X}}(x) F_{\rm{N}}(y-x) dx}.
\end{equation}
An alternative expression can be obtained by exchanging the integration order, which leads to
\begin{equation}
\label{eq:theta_i_noise_2}
\theta_i = \int_{-\infty}^{\infty}{f_{\rm{N}}(n)  \int_{y_{i-1}-n}^{y_i-n} {x
f_{\rm{X}}(x) dx} dn} = D(y_i) - D(y_{i-1}),
\end{equation}
where
\begin{align}
\label{eq:d_2}
D(y) =& \int_{-\infty}^{\infty} { \scalebox{0.9}{$ f_{\rm{N}}(n)  \left[ (y-n)
F_{\rm{X}}(y-n) - I_{\rm{X}}(y-n) \right] $} dn}, \\ I_{\rm{X}}(y) =&
\int_{-\infty}^{y}{F_{\rm{X}}(x)dx}.
\end{align}
Which expression is preferable, between \eqref{eq:d_1} and \eqref{eq:d_2}, depends on the expressions of $f_{\rm{X}}(x)$ and $f_{\rm{N}}(n)$.

Using \eqref{eq:def_phi_ij_1} and \eqref{eq:noise_model}, we obtain
\begin{align}
\label{eq:phi_ii_noise_1}
R_{ii} =& F_{\rm{Y}}(y_i) - F_{\rm{Y}}(y_{i-1}), \\
\label{eq:f_1}
F_{\rm{Y}}(y) =& \int_{-\infty}^{\infty}{f_{\rm{X}}(x) F_{\rm{N}}(y-x) dx} \\
\label{eq:f_2}
          =& \int_{-\infty}^{\infty}{f_{\rm{N}}(n) F_{\rm{X}}(y-x) dn}.
\end{align}
Thus, using \eqref{eq:g_i_mmse}, either \eqref{eq:theta_i_noise_1} or \eqref{eq:theta_i_noise_2}, and \eqref{eq:phi_ii_noise_1},
the Q-MMSE estimator for the additive noise model $\eqref{eq:noise_model}$ is expressed by
\begin{equation}
\label{eq:g_i_QMMSE_add_model}
g_{i,\textrm{Q-MMSE}} = \frac{ D(y_i) - D(y_{i-1}) }{F(y_i) - F(y_{i-1})}.
\end{equation}

\section{A Numerical Example}
In this section, we want to numerically compare the MSE and the SNR performances of the Q-MMSE estimator with those of the optimal MMSE estimator, in order to show the usefulness of Q-MMSE estimators with a limited number of quantization levels. Therefore, first we derive the mathematical expressions of the optimal MMSE estimator and of the Q-MMSE estimator, assuming a non-trivial additive noise model \eqref{eq:noise_model} where both the signal and the noise are non-Gaussian. Specifically, we model the signal $x$ with a Laplace pdf
\begin{equation}
\label{eq:pdf_signal_laplace}
f_{\rm{X}}(x) = \frac{\alpha}{2} e^{-\alpha |x|},
\end{equation}
with $\alpha = \sqrt{2}/\sigma_x$, and the noise $n$ with a Laplace mixture pdf 
\begin{equation}
\label{eq:pdf_noise_bernoulli_laplace}
f_{\rm{N}}(n) = p_0 \frac{\beta_0}{2} e^{-\beta_0 |n|} + p_1 \frac{\beta_1}{2}
e^{-\beta_1 |n|},
\end{equation}
with $\{\beta_m = \sqrt{2}/\sigma_{n,m}\}_{m=0,1}$, $R =
\sigma^2_{n,0} / \sigma^2_{n,1}$, $\sigma^2_n = p_0 \sigma^2_{n,0} + p_1
\sigma^2_{n,1}$, $p_0 + p_1 = 1$ and $0 \le p_0 \le 1$.
Basically, \eqref{eq:pdf_noise_bernoulli_laplace} models a noise generated by two independent sources: each noise source, characterized by a Laplace pdf with average power $\sigma^2_{n,m}$, occurs with probability $p_m$. Similar results can be obtained by modeling either the noise, or the signal, or both, as a Gaussian mixture, thus covering a wide range of practical applications of non-Gaussian denoising.

As detailed in Appendix~B, direct computation of \eqref{eq:g_mmse} with \eqref{eq:pdf_signal_laplace} and \eqref{eq:pdf_noise_bernoulli_laplace} yields the optimal MMSE estimator
\begin{align}
\label{eq:g_mmse_laplace}
& g_{\textrm{MMSE}}(y) = \sgn{(y)} \times \nonumber \\
& \times  \frac{ \sum\limits_{m=0}^{1} p_m
\left[ C_{1,m} (e^{-\beta_m |y|} - e^{-\alpha |y|}) - C_{2,m} \beta_m |y| e^{-\alpha |y|} \right]}
{\sum\limits_{m=0}^{1} { p_m C_{2,m} \left( \alpha e^{-\beta_m |y|} - \beta_m e^{-\alpha |y|} \right) }}, \\
\label{eq:C_1m_and_C_2m}
& C_{1,m} = \frac{\alpha^2 \beta_m^2}{(\alpha^2 - \beta_m^2)^2}, \qquad C_{2,m} = \frac{\alpha \beta_m}{2(\alpha^2 - \beta_m^2)}.
\end{align}

The Q-MMSE estimator can be calculated by solving \eqref{eq:d_1} and \eqref{eq:f_1} using the pdf in \eqref{eq:pdf_signal_laplace} and \eqref{eq:pdf_noise_bernoulli_laplace}: as detailed in Appendix~C, when $y>0$, this calculation leads to
\begin{align}
\label{eq:d_mmse_laplace}
D(y) = &  \sum\limits_{m=0}^{1} p_m \frac{ \beta_m^2 (3 \alpha^2  - \beta_m^2) e^{-\alpha y} }{2 \alpha (\alpha^2 - \beta_m^2)^2}  \nonumber \\
     - &  \sum\limits_{m=0}^{1} p_m{\frac{ \alpha^2  \beta_m e^{-\beta_m y} } { (\alpha^2 - \beta_m^2)^2} }  +
     \sum\limits_{m=0}^{1} p_m \frac{ \beta_m^2  y  e^{-\alpha  y} }{2    (\alpha^2 - \beta_m^2)  }, \\
\label{eq:f_mmse_laplace}
F_{\rm{Y}}(y) = & 1 - \sum\limits_{m=0}^{1}  {p_m \frac{ \alpha^2  e^{-\beta_m y} - \beta_m^2 e^{-\alpha y}}{2 (\alpha^2 - \beta_m^2)} },
\end{align}
which inserted into \eqref{eq:g_i_QMMSE_add_model} give the final result.

In addition to MMSE and Q-MMSE, other two alternative estimators are included in this comparison: (a) the sampled MMSE (S-MMSE) estimator $g_{i,\textrm{S-MMSE}}$, obtained by sampling the optimal MMSE estimator $g_{\textrm{MMSE}}(\cdot)$ at the midpoint of each quantization interval, e.g., $ g_{i,\textrm{S-MMSE}} = g_{\textrm{MMSE}}((y_{i-1}+y_{i})/2) $; and (b) the optimal quantizer (OQ) obtained by applying the Lloyd-Max algorithm \cite{Gersho:77} to the signal pdf $f_{\rm{X}}(x)$. Note that the Lloyd-Max OQ exploits the statistical knowledge of the parameter of interest $x$ only, and neglects the noise, while the Q-MMSE estimator-quantizer also exploits the knowledge of the pdf of noise $n$: hence, the Q-MMSE estimator is expected to give better performance.

With reference to the choice of the $N-1$ thresholds $\{y_i\}$ of the Q-MMSE estimators, a heuristic approach chooses all the $N-1$ thresholds equispaced, such that the overload probability $P_{\textrm{ol}}=P\{y \in [-\infty,y_1)\cup[y_{N-1},\infty)\}$ of the quantizer is fixed: this limits the amount of saturating distortion. Another option is to choose the non-uniform thresholds $\{y_i\}$ given by the Lloyd-Max algorithm \cite{Gersho:77} applied to the signal pdf $f_{\rm{X}}(x)$ in \eqref{eq:pdf_signal_laplace}. For all the quantized estimators, we use the acronym NU for non-uniform quantization and U for uniform quantization.

\begin{figure} [t]
\centerline{\includegraphics[width=1\textwidth]{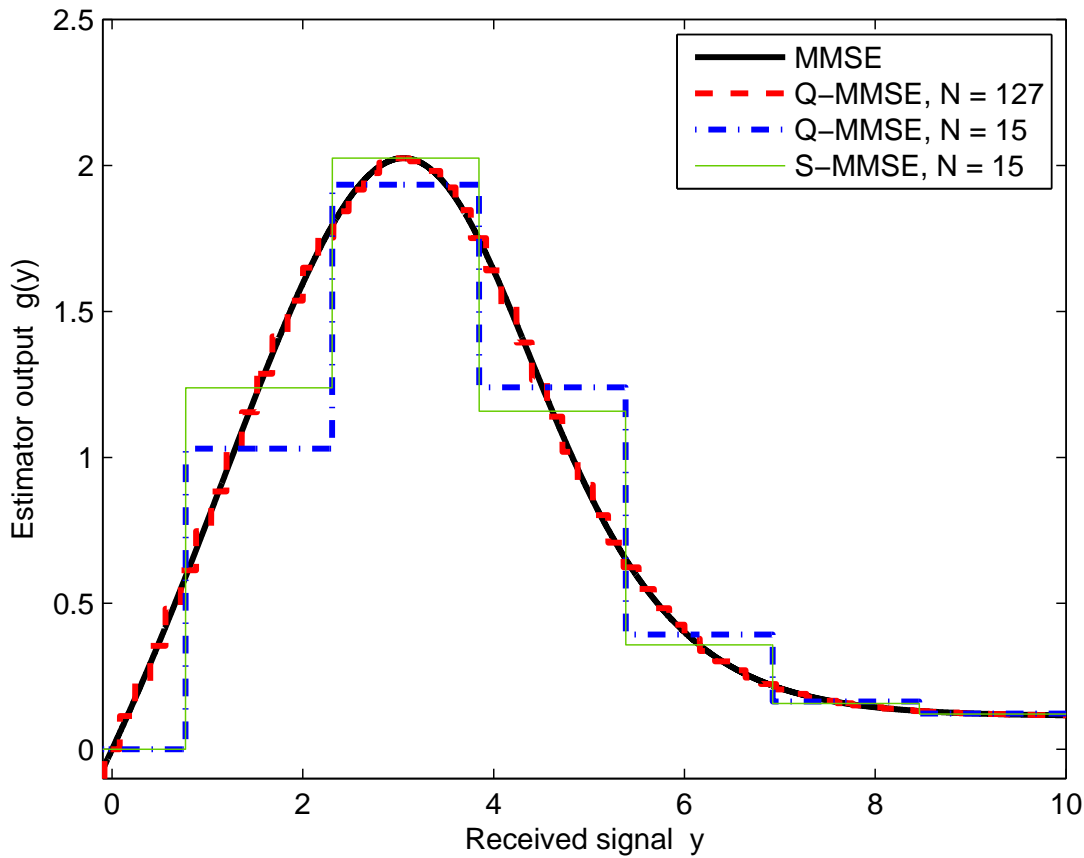}}
\caption{Comparison between the optimal (unquantized) MMSE estimator and Q-MMSE estimators with uniform quantization ($N$ is the number of intervals).}
\label{fig1}
\end{figure}

Fig.~\ref{fig1} compares the shape of the Q-MMSE estimator $g_{\textrm{Q-MMSE}}(\cdot)$ with the shape of the optimal (unquantized) MMSE estimator $g_{\textrm{MMSE}}(\cdot)$, when $\sigma_x = 1$, $\sigma_n = 4$, $R = 0.001$, $p_0 = 0.9$, and the $N-1$ thresholds are equispaced between $y_1=-10$ and $y_{N-1}=10$, which induce an overload probability $P_{\textrm{ol}} \approx 0.0327$. Since all the considered MMSE estimators are odd functions of the input $y$, Fig.~\ref{fig1} only displays the positive half. Fig.~\ref{fig1} confirms that, when the number $N$ of quantization levels increases, the Q-MMSE estimator tends to the optimal MMSE estimator. Note also that the Q-MMSE estimator $g_{i,\textrm{Q-MMSE}}$ is different from the staircase curve of the S-MMSE estimator $g_{i,\textrm{S-MMSE}}$.

\begin{figure} [t]
\centerline{\includegraphics[width=1\textwidth]{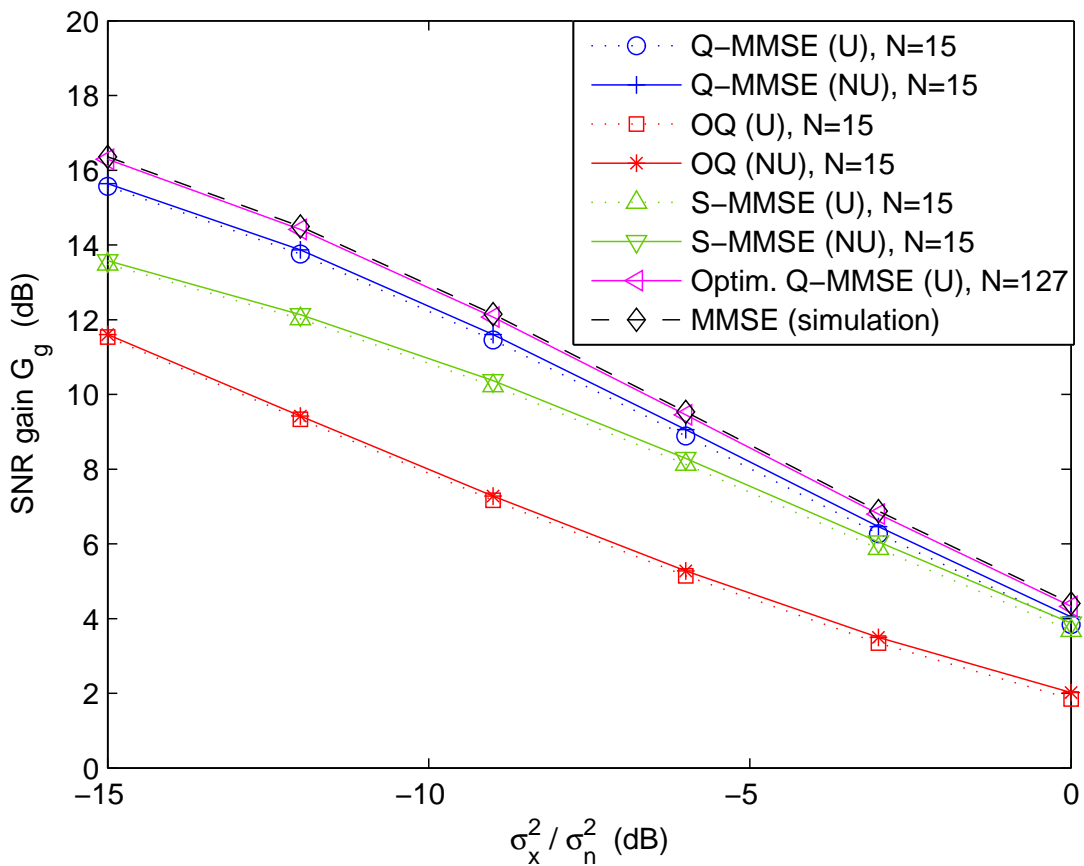}}
\caption{SNR gain $G_g$ of different estimators as a function of the input SNR $\sigma_x^2 / \sigma_n^2$.}
\label{fig2}
\end{figure}

\begin{figure} [t]
\centerline{\includegraphics[width=1\textwidth]{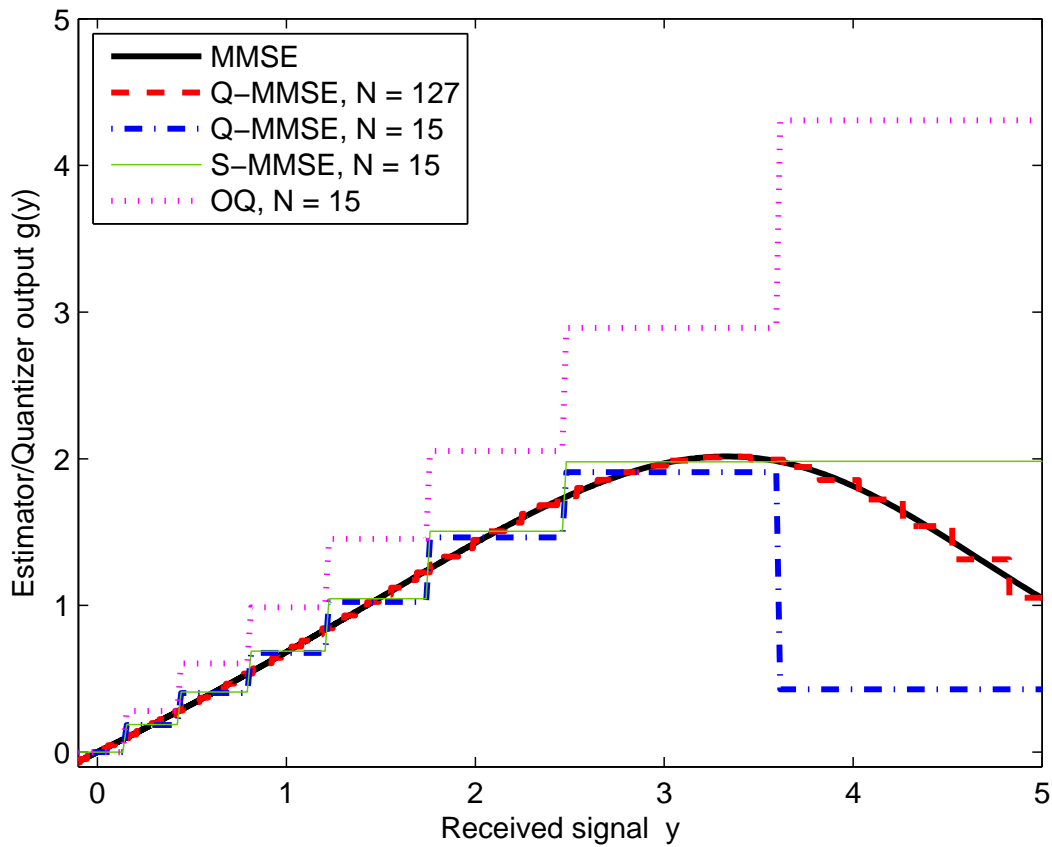}}
\caption{Comparison between different estimators with non-uniform quantization
($N$ is the number of intervals).}
\label{fig3}
\end{figure}

\begin{figure} [t]
\centerline{\includegraphics[width=1\textwidth]{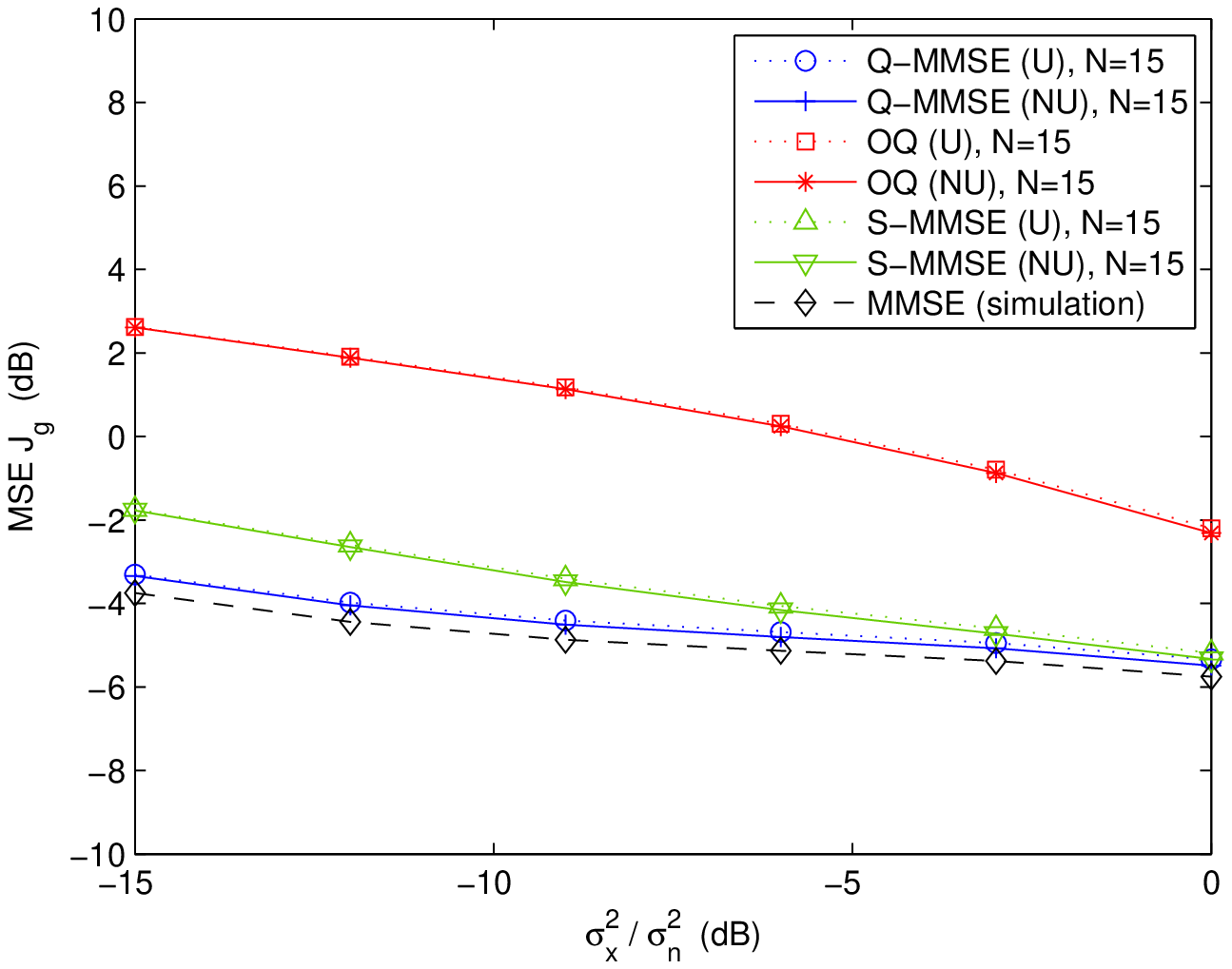}}
\caption{MSE $J_g$ of different estimators as a function of the input SNR $\sigma_x^2 / \sigma_n^2$.}
\label{fig4}
\end{figure}

Fig.~\ref{fig2} shows the SNR gain $G_g$ provided by different estimators $g(\cdot)$. The SNR
gain $G_g$ is defined as
\begin{equation}
\label{eq:gain}
G_g = \frac{\gamma_g}{\sigma_x^2 / \sigma_n^2},
\end{equation}
where $\gamma_g$ is the SNR at the output of the estimator, and $\sigma_x^2 / \sigma_n^2$ is the SNR at the input of the estimator. The signal and noise parameters are the same of Fig.~\ref{fig1}, except for the variable $\sigma_n$. Fig.~\ref{fig2} compares the SNR performance of Q-MMSE, S-MMSE, and OQ estimators, assuming uniform and non-uniform quantization versions (with labels U and NU in the legend of Fig.~\ref{fig2}): the overload regions are the same for both versions and have been selected by the Lloyd-Max algorithm, which ends up with an overload probability $P_{\textrm{ol}} \approx 0.0093$ when $\sigma_x^2 / \sigma_n^2 = 0$~dB and $P_{\textrm{ol}} \approx 0.0805$ when $\sigma_x^2 / \sigma_n^2 = -12$~dB. As a reference, Fig.~\ref{fig2} also includes an optimal Q-MMSE (with $N=127$) with uniform quantization obtained by an exhaustive maximization of the SNR gain over all the possible choices for the overload regions (i.e., for all the possible choices of $y_1=-y_{N-1}$): this is equivalent to an optimization of the interval size $\Delta y = (y_{N-1}-y_1)/(N-2)$ of the uniform quantization intervals. When the number of quantization intervals $N$ is sufficiently high, the SNR of this optimal Q-MMSE estimator basically coincides with the SNR of the optimal (unquantized) MMSE, whose simulated SNR gain is included in Fig.~\ref{fig2} as well.

Fig.~\ref{fig2} confirms that the SNR gain of the Q-MMSE estimator is larger than for the other quantized estimators, provided that the quantization intervals are the same. The SNR of the Q-MMSE estimator can be further improved by increasing the number of intervals and by optimizing the (uniform) interval sizes, as shown in Fig.~\ref{fig2} by the curve with $N = 127$ with optimized overload regions. In addition, Fig.~\ref{fig2} shows that the SNR of the optimal Q-MMSE estimator is very close to the simulated SNR of the optimal (unquantized) MMSE estimator. Therefore, the proposed Q-MMSE approach permits to obtain analytical tight lower bounds on the SNR of the optimal (unquantized) MMSE estimator.

Fig.~\ref{fig3} compares the function $g(y)$ for the estimators of Fig.~\ref{fig2} with non-uniform quantization, when $\sigma_x^2 / \sigma_n^2 =-15$ dB. Fig.~\ref{fig3} highlightsthat the function $g(y)$ of the Lloyd-Max OQ is nondecreasing, because the noise is neglected; differently, the function $g(y)$ of the (Q-) MMSE estimators can be non-monotonic, like in this specific example.

Fig.~\ref{fig4} displays the MSE of different estimators, in the same scenario of Fig.~\ref{fig2}. It is evident that the Q-MMSE estimator provides the lowest MSE among all the quantized estimators that use the same quantization intervals. Note that the analytical MSE of the Q-MMSE estimator can be used as an upper bound of the minimum value $J_\textrm{MMSE}$ (obtained in Fig.~\ref{fig4} by simulation). Similarly to the SNR analysis of Fig.~\ref{fig2}, tighter upper bounds on the MMSE $J_{g_\textrm{MMSE}}$ can be obtained by increasing the number of intervals $N$ and by further optimization over all the possible overload regions.

\section{Conclusion}
In this paper, we have studied a meaningful definition of the MSNR estimator, and we established its equivalence with the MMSE estimator, regardless of the statistics of the noise and of the parameter of interest. We have also extended this equivalence to a specific class of suboptimal estimators expressed as a linear combination of arbitrary (fixed) functions; conversely, we have explained that the same equivalence does not hold true in general for non-BEM suboptimal estimators.

The developed theoretical framework has been instrumental to study Bayesian estimators whose input is a quantized observation of a parameter of interest corrupted by an additive noise. We have shown that, when the size of the quantization intervals goes to zero, the Q-MMSE (Q-MSNR) estimator exactly tends to the MMSE (MSNR) estimator for unquantized observations. Furthermore, by a practical example, we have shown that, using a fairly limited number of quantization levels, the Q-MMSE estimator can easily approach the performance of the optimal (unquantized) MMSE estimator: the designed Q-MMSE estimator, clearly, outperforms in SNR (and in MSE) other suboptimal estimators.

\section*{Appendix A - Other BEM-Based Estimators}
We detail BEM-based estimators that produce the maximum SNR, similarly to B-MMSE and B-MSNR estimators: unbiased estimators and a maximum-gain estimator.

Unbiased estimators are defined by $E_{\rm{Y|X}} \{ g(y)|x \} =x $ and hence are
characterized by $K_{g} = 1$ in \eqref{eq:lin_regr}. By \eqref{eq:def_k},
\eqref{eq:g_bem}, \eqref{eq:def_g}--\eqref{eq:def_theta_i}, for the BEM-based estimators we have
\begin{equation}
\label{eq:k_1}
K_{g} = \frac{\mathbf{g}^T\boldsymbol{\theta}}{\sigma^2_x}.
\end{equation}
Therefore, the BEM-based unbiased MSNR (B-UMSNR) estimator is obtained by maximizing \eqref{eq:snr_1} subject to the constraint $\mathbf{g}^T\boldsymbol{\theta} = \sigma^2_x$, while the BEM-based unbiased MMSE (B-UMMSE) estimator is obtained by minimizing \eqref{eq:mse_2} subject to the same constraint. By inserting the constraint $\mathbf{g}^T\boldsymbol{\theta} = \sigma^2_x$ into \eqref{eq:snr_1} and \eqref{eq:mse_2}, both optimizations are equivalent to the minimization of the output power $E_{\rm{Y}}\{g^2(y)\} = \mathbf{g}^T \mathbf{R} \mathbf{g}$ subject to $\mathbf{g}^T\boldsymbol{\theta} = \sigma^2_x$, which leads to
\begin{equation}
\label{eq:unbiased}
\mathbf{g}_{\textrm{B-UMSNR}} = \mathbf{g}_{\textrm{B-UMMSE}} = \frac{\sigma^2_x
}{\boldsymbol{\theta}^T \mathbf{R}^{-1} \boldsymbol{\theta}}
\mathbf{R}^{-1} \boldsymbol{\theta}.
\end{equation}
The solution \eqref{eq:unbiased} is equivalent to \eqref{eq:g_msnr_1} with
\begin{equation}
\label{c_unbiased}
c = \frac{\sigma^2_x (\sigma^2_x - \boldsymbol{\theta}^T \mathbf{R}^{-1}
\boldsymbol{\theta})}{\boldsymbol{\theta}^T \mathbf{R}^{-1}
\boldsymbol{\theta}}.
\end{equation}
Hence, the B-UMMSE estimator gives the maximum SNR achievable by BEM-based estimators, and is a scaled version
of the B-MMSE estimator \eqref{eq:g_mmse_constr}.

An alternative Bayesian criterion is the maximization of the gain $K_{g}$ \eqref{eq:def_k} or \eqref{eq:k_1}, subject to a power constraint. Using the output power constraint $E_{\rm{Y}}\{g^2(y)\} = \mathbf{g}^T \mathbf{R} \mathbf{g} = P$, the BEM-based maximum-gain (B-MG) estimator is expressed by
\begin{equation}
\label{eq:opcmg}
\mathbf{g}_{\textrm{B-MG}} = \sqrt{\frac{P}{\boldsymbol{\theta}^T
\mathbf{R}^{-1} \boldsymbol{\theta}}} \mathbf{R}^{-1}
\boldsymbol{\theta},
\end{equation}
which is a scaled version of the B-MMSE estimator and hence an MSNR estimator among the BEM-based estimators.

\section*{Appendix B - Derivation of \eqref{eq:g_mmse_laplace}}
Here we show that the computation of \eqref{eq:g_mmse}, when the signal pdf is \eqref{eq:pdf_signal_laplace} and the noise pdf is \eqref{eq:pdf_noise_bernoulli_laplace}, leads to \eqref{eq:g_mmse_laplace}. First, using the Bayes' theorem, the MMSE estimator \eqref{eq:g_mmse} is rewritten as
\begin{equation}
\label{eq:g_mmse_2}
g_{\textrm{MMSE}}(y) = \frac{\int_{-\infty}^{\infty}{x f_{\rm{Y|X}}(y|x) f_{\rm{X}}(x) dx}}{f_{\rm{Y}}(y)};
\end{equation}
in addition, the noise pdf \eqref{eq:pdf_noise_bernoulli_laplace} can be rewritten as
\begin{align}
\label{eq:pdf_noise_bernoulli_laplace_2}
f_{\rm{N}}(n) =& \sum\limits_{m=0}^{1} p_m f_{{\rm{N}},m}(n) \\
\label{eq:pdf_noise_bernoulli_laplace_3}
f_{{\rm{N}},m}(n) =& \frac{\beta_m}{2} e^{-\beta_m |n|}.
\end{align}
Using \eqref{eq:noise_model}, \eqref{eq:pdf_signal_laplace}, \eqref{eq:pdf_noise_bernoulli_laplace},  \eqref{eq:pdf_noise_bernoulli_laplace_2}, and \eqref{eq:pdf_noise_bernoulli_laplace_3}, the numerator of \eqref{eq:g_mmse_2}, for $y>0$, can be rewritten as
\begin{align}
\label{eq:num_mmse_2}
  &                           \int_{-\infty}^{\infty}{x f_{\rm{Y|X}}(y|x)      f_{\rm{X}}(x) dx} \\
= &                           \int_{-\infty}^{\infty}{x f_{\rm{N}}(y - x)      f_{\rm{X}}(x) dx} \\
= & \sum\limits_{m=0}^{1} p_m \int_{-\infty}^{\infty}{x f_{{\rm{N}},m} (y - x) f_{\rm{X}}(x) dx} \\
\label{eq:num_mmse_2_end}
= & I_1(y) + I_2(y) + I_3(y),
\end{align}
where
\begin{align}
\label{eq:num_first}
I_1(y) = & \sum\limits_{m=0}^{1} p_m \frac{\alpha \beta_m}{4} e^{- \beta_m y} \int_{-\infty}^{0}{ x  e^{( \alpha + \beta_m) x} dx}, \\
\label{eq:num_second}
I_2(y) = & \sum\limits_{m=0}^{1} p_m \frac{\alpha \beta_m}{4} e^{- \beta_m y} \int_{0}^{y}{       x  e^{(-\alpha + \beta_m) x} dx}, \\
\label{eq:num_third}
I_3(y) = & \sum\limits_{m=0}^{1} p_m \frac{\alpha \beta_m}{4} e^{  \beta_m y} \int_{y}^{\infty}{  x  e^{(-\alpha - \beta_m) x} dx}.
\end{align}
The three integrals \eqref{eq:num_first}, \eqref{eq:num_second}, and \eqref{eq:num_third}, can be solved using
\begin{equation}
\label{eq:primitive}
\int{ x  e^{a x} dx} = \begin{cases} \frac{1}{a^2}[(ax-1) e^{a x}] + C, & \mbox{if } a \ne 0,\\
\frac{1}{2}x^2 + C, & \mbox{if } a = 0, \end{cases}
\end{equation}
where $C$ is an arbitrary constant. If we assume that $\alpha \ne \beta_m$, for $m=0,1$, then \eqref{eq:num_first}--\eqref{eq:num_third} become
\begin{align}
\label{eq:num_first_2}
I_1(y) = & - \sum\limits_{m=0}^{1} p_m \frac{\alpha \beta_m e^{- \beta_m y}}{4(\alpha + \beta_m)^2}, \\
\label{eq:num_second_2}
I_2(y) = &   \sum\limits_{m=0}^{1} p_m \frac{\alpha \beta_m \{ [ (\beta_m - \alpha) y -1 ] e^{- \alpha y}  + e^{- \beta_m y} \}} {4(\beta_m - \alpha)^2}, \\
\label{eq:num_third_2}
I_3(y) = &   \sum\limits_{m=0}^{1} p_m \frac{\alpha \beta_m [ (\alpha + \beta_m) y +1 ] e^{- \alpha y} }{4(\alpha + \beta_m)^2}  .
\end{align}
Hence, the numerator of \eqref{eq:g_mmse_2}, for $y>0$, is equal to
\begin{align}
& I_1(y) + I_2(y) + I_3(y) = \\
\label{eq:num_mmse_3}
& = \sum\limits_{m=0}^{1} p_m  \left[ C_{1,m} (e^{-\beta_m y} - e^{-\alpha y}) - C_{2,m} \beta_m y e^{-\alpha y} \right],
\end{align}
where $C_{1,m}$ and $C_{2,m}$ are expressed by \eqref{eq:C_1m_and_C_2m}. If we repeat the same procedure for $y<0$, we obtain a similar equation.

On the other hand, using \eqref{eq:noise_model}, \eqref{eq:pdf_signal_laplace}, \eqref{eq:pdf_noise_bernoulli_laplace_2} and \eqref{eq:pdf_noise_bernoulli_laplace_3}, the denominator of \eqref{eq:g_mmse_2} is equal to
\begin{align}
\label{eq:den_mmse_2}
f_{\rm{Y}}(y) = & f_{\rm{X}}(y) \ast f_{\rm{N}}(y) = f_{\rm{X}}(y) \ast \sum\limits_{m=0}^{1} p_m f_{{\rm{N}},m}(y) \\
\label{eq:den_mmse_3}
             = & \sum\limits_{m=0}^{1} p_m [ f_{\rm{X}}(y) \ast f_{{\rm{N}},m}(y) ] =  \sum\limits_{m=0}^{1} p_m f_{{\rm{Y}},m}(y),
\end{align}
where $\ast$ denotes convolution and
\begin{equation}
\label{eq:convolution_laplace}
f_{{\rm{Y}},m}(y) = f_{\rm{X}}(y) \ast f_{{\rm{N}},m}(y).
\end{equation}
By denoting with $C_{\rm{X}}(u)$ the characteristic function associated with the pdf $f_{\rm{X}}(x)$, \eqref{eq:convolution_laplace} translates into
\begin{equation}
\label{eq:characteristic_convolution_laplace}
C_{{\rm{Y}},m}(u) = C_{\rm{X}}(u) C_{{\rm{N}},m}(u) = \frac{\alpha^2}{\alpha^2 + 4 \pi^2 u^2} \frac{\beta_m^2}{\beta_m^2 + 4 \pi^2 u^2}.
\end{equation}
If we assume that $\alpha \ne \beta_m$, for $m=0,1$, then \eqref{eq:characteristic_convolution_laplace} can be decomposed in partial fractions as
\begin{align}
\label{eq:characteristic_convolution_laplace_2}
C_{{\rm{Y}},m}(u) =& \frac{\beta_m^2}{\beta_m^2 - \alpha^2} \frac{\alpha^2}{\alpha^2 + 4 \pi^2 u^2} + \frac{\alpha^2}{\alpha^2 - \beta_m^2} \frac{\beta_m^2}{\beta_m^2 + 4 \pi^2 u^2}\\
                  =& \frac{\beta_m^2}{\beta_m^2 - \alpha^2} C_{\rm{X}}(u) + \frac{\alpha^2}{\alpha^2 - \beta_m^2} C_{{\rm{N}},m}(u),
\end{align}
which, by means of \eqref{eq:den_mmse_2} and \eqref{eq:den_mmse_3}, leads to
\begin{align}
\label{eq:convolution_laplace_2}
& f_{{\rm{Y}},m}(y) = \frac{\beta_m^2}{\beta_m^2 - \alpha^2} f_{\rm{X}}(y) + \frac{\alpha^2}{\alpha^2 - \beta_m^2} f_{{\rm{N}},m}(y),\\
\label{eq:convolution_laplace_3}
& f_{{\rm{Y}}}(y)   = \sum\limits_{m=0}^{1} p_m \left[ \frac{\beta_m^2}{\beta_m^2 - \alpha^2} f_{\rm{X}}(y) + \frac{\alpha^2}{\alpha^2 - \beta_m^2} f_{{\rm{N}},m}(y) \right].
\end{align}
Therefore, by \eqref{eq:convolution_laplace_3}, \eqref{eq:pdf_signal_laplace}, \eqref{eq:pdf_noise_bernoulli_laplace}, and \eqref{eq:pdf_noise_bernoulli_laplace_2}, the denominator of \eqref{eq:g_mmse_2} is equal to
\begin{align}
\label{eq:den_mmse_4}
f_{{\rm{Y}}}(y) =& \sum\limits_{m=0}^{1} p_m \frac{\alpha \beta_m^2 e^{-\alpha |y|}}{2(\beta_m^2 - \alpha^2)}  + \sum\limits_{m=0}^{1} p_m  \frac{\alpha^2 \beta_m e^{-\beta_m |y|}}{2(\alpha^2 - \beta_m^2)} \\
\label{eq:den_mmse_5}
=& \sum\limits_{m=0}^{1} p_m C_{2,m} (\alpha e^{-\beta_m |y|} - \beta_m e^{-\alpha |y|}),
\end{align}
where $C_{2,m}$ is expressed by \eqref{eq:C_1m_and_C_2m}. By inserting \eqref{eq:num_mmse_2}--\eqref{eq:num_mmse_2_end}, \eqref{eq:num_mmse_3}, and \eqref{eq:den_mmse_4}--\eqref{eq:den_mmse_5} into \eqref{eq:g_mmse_2}, we obtain the mathematical expression of $g_{\textrm{MMSE}}(y)$ for $y>0$, and, by repeating the same procedure for negative values of $y$, we obtain the final expression of $g_{\textrm{MMSE}}(y)$ reported in  \eqref{eq:g_mmse_laplace}--\eqref{eq:C_1m_and_C_2m}, which is valid for all values of $y$. Note that, since the signal pdf and the noise pdf are both symmetric, the MMSE estimator is and odd function of $y$, and therefore $g_{\textrm{MMSE}}(-y) = -g_{\textrm{MMSE}}(y)$.

\section*{Appendix C - Derivation of \eqref{eq:d_mmse_laplace} and \eqref{eq:f_mmse_laplace}}
Herein we detail the computation of $D(y)$ in \eqref{eq:d_mmse_laplace} and of $F_{\rm{Y}}(y)$ in \eqref{eq:f_mmse_laplace}: these two quantities are derived by calculating \eqref{eq:d_1} and \eqref{eq:f_1}, respectively, for the additive noise model \eqref{eq:noise_model}, when the signal pdf is expressed by \eqref{eq:pdf_signal_laplace} and the noise pdf is expressed by \eqref{eq:pdf_noise_bernoulli_laplace}. Indeed, \eqref{eq:d_1} and \eqref{eq:f_1} are necessary in order to compute the Q-MMSE estimator, expressed by \eqref{eq:g_i_mmse}, via \eqref{eq:theta_i_noise_1}--\eqref{eq:d_1} and \eqref{eq:phi_ii_noise_1}--\eqref{eq:f_2}. The derivations of $D(y)$ and $F_{\rm{Y}}(y)$ are performed only for $y>0$ (those for $y<0$ are similar).

By \eqref{eq:pdf_noise_bernoulli_laplace_2} and \eqref{eq:pdf_noise_bernoulli_laplace_3}, the noise cdf can be expressed as
\begin{align}
\label{eq:cdf_noise_bernoulli_laplace}
F_{\rm{N}}(n) =& \sum\limits_{m=0}^{1} p_m F_{{\rm{N}},m}(n) \\
\label{eq:cdf_noise_bernoulli_laplace_2}
F_{{\rm{N}},m}(n) =&
\begin{cases} \frac{1}{2}e^{\beta_m n}, & \mbox{if } n<0,\\
1 - \frac{1}{2}e^{-\beta_m n}, & \mbox{if } n \ge 0, \end{cases}
\end{align}
and therefore, by \eqref{eq:pdf_signal_laplace}, $D(y)$ in \eqref{eq:d_1} becomes
\begin{align}
\label{eq:d_mmse_laplace_1}
D(y) =& \sum\limits_{m=0}^{1} p_m                                 \int_{-\infty}^{\infty}{x f_{\rm{X}}(x) F_{{\rm{N}},m}(y-x) dx} \\
     =& I_4(y) + I_5(y) + I_6(y) + I_7(y) + I_8(y),
\end{align}
where
\begin{align}
\label{eq:I_4}
I_4(y) =&   \sum\limits_{m=0}^{1} p_m \frac{\alpha}{2}                \int_{-\infty}^{     0}{x e^{ \alpha x} dx}, \\
I_5(y) =& - \sum\limits_{m=0}^{1} p_m \frac{\alpha}{4} e^{-\beta_m y} \int_{-\infty}^{     0}{x e^{ (\alpha + \beta_m ) x} dx}, \\
I_6(y) =&   \sum\limits_{m=0}^{1} p_m \frac{\alpha}{2}                \int_{      0}^{     y}{x e^{-\alpha x} dx}, \\
I_7(y) =& - \sum\limits_{m=0}^{1} p_m \frac{\alpha}{4} e^{-\beta_m y} \int_{      0}^{     y}{x e^{(-\alpha + \beta_m ) x} dx}, \\
\label{eq:I_8}
I_8(y) =&   \sum\limits_{m=0}^{1} p_m \frac{\alpha}{4} e^{ \beta_m y} \int_{      y}^{\infty}{x e^{-(\alpha + \beta_m ) x} dx}.
\end{align}
By assuming $\alpha \ne \beta_m$, for $m=0,1$, and by solving the five integrals in \eqref{eq:I_4}--\eqref{eq:I_8} using \eqref{eq:primitive}, it is easy to show that $D(y)$ in \eqref{eq:d_mmse_laplace_1} becomes equal to \eqref{eq:d_mmse_laplace}.

The cdf $F_{\rm{Y}}(y)$ can be easily calculated from \eqref{eq:convolution_laplace_2} and \eqref{eq:cdf_noise_bernoulli_laplace_2}, which lead to
\begin{align}
\label{eq:cdf_y_laplace_2}
F_{{\rm{Y}},m}(y) =& \frac{\beta_m^2}{\beta_m^2 - \alpha^2} F_{\rm{X}}(y) + \frac{\alpha^2}{\alpha^2 - \beta_m^2} F_{{\rm{N}},m}(y)\\
\label{eq:cdf_y_laplace_3}
                  =& 1 + \frac{\beta_m^2 e^{-\alpha y} - \alpha^2 e^{-\beta_m y}}{2(\alpha^2 - \beta_m^2)},
\end{align}
where we have used $F_{\rm{X}}(y) = 1 - \frac{1}{2} e^{-\alpha y}$ for $y>0$. Using \eqref{eq:cdf_y_laplace_3} with
 \eqref{eq:den_mmse_2}--\eqref{eq:den_mmse_3}, we obtain the final expression \eqref{eq:f_mmse_laplace}.


\begin{thebibliography}{12}
\bibitem{Kay:93} S. M. Kay, \em Fundamentals of Statistical Signal Processing: Estimation Theory\em. Englewood Cliffs, NJ: Prentice-Hall, 1993.
\bibitem{Poor:94} H. V. Poor, \em An Introduction to Signal Detection and Estimation\em, 2nd ed. New York, NY: Springer-Verlag, 1994.
\bibitem{Guo:05} D. Guo, S. Shamai, and S. Verd\'u, ``Mutual information and minimum mean-square error in Gaussian channels,'' \em IEEE Trans. Inf. Theory\em, vol.~51, no.~4, pp.~1261--1283, Apr.~2005.
\bibitem{Guo:08} D. Guo, S. Shamai, and S. Verd\'u, ``Mutual information and conditional mean estimation in Poisson channels,'' \em IEEE Trans. Inf. Theory\em, vol.~54, no.~5, pp.~1837--1849, May~2008.
\bibitem{Guo:11} D. Guo, Y. Wu, S. Shamai, and S. Verd\'u, ``Estimation in Gaussian noise: Properties of the minimum mean-square error,'' \em IEEE Trans. Inf. Theory\em, vol.~57, no.~4, pp.~2371--2385, Apr.~2011.
\bibitem{Venkat:12} K. Venkat and T. Weissman, ``Pointwise relations between information and estimation in Gaussian noise,'' \em IEEE Trans. Inf. Theory\em, vol.~58, no.~10, pp.~6264-6281, Oct.~2012.
\bibitem{Wu:12} Y. Wu and S. Verd\'u, ``Functional properties of minimum mean-square error and mutual information,'' \em IEEE Trans. Inf. Theory\em, vol.~58, no.~3, pp.~1289--1301, Mar.~2012.
\bibitem{Yang:07} Y. Yang and R. S. Blum, ``MIMO radar waveform design based on mutual information and minimum mean-square error estimation,'' \em IEEE Trans. Aerosp. Electron. Syst.\em, vol.~43, no.~1, pp.~330--343, Jan.~2007.
\bibitem{Gardner:80} W. Gardner, ``A unifying view of second-order measures of quality for signal classification,'' \em IEEE Trans. Commun.\em, vol.~28, no.~6, pp.~807--816, June~1980.
\bibitem{Zhidkov:06} S. V. Zhidkov, ``Performance analysis and optimization of OFDM receiver with blanking nonlinearity in impulsive noise environment,'' \em IEEE Trans. Veh. Technol.\em, vol.~55, no.~1, pp.~234--242, Jan.~2006.
\bibitem{Kay:98} S. M. Kay, \em Fundamentals of Statistical Signal Processing: Detection Theory\em. Englewood Cliffs, NJ: Prentice-Hall, 1998.
\bibitem{Akyol:12} E. Akyol, K. B. Viswanatha, and K. Rose, ``On conditions for linearity of optimal estimation,'' \em IEEE Trans. Inf. Theory\em, vol.~58, no.~6, pp.~3497--3508, June~2012.
\bibitem{Giannakis:98} G. B. Giannakis and C. Tepedelenlio{\u{g}}lu, ``Basis expansion models and diversity techniques for blind identification and equalization of time-varying channels,'' \em Proc. of the IEEE\em, vol.~86, no.~10, pp.~1969--1986, Oct.~1998.
\bibitem{Flam:12} J. T. Fl{\aa}m, S. Chatterjee, K. Kansanen, and T. Ekman, ``On MMSE estimation: A linear model under Gaussian mixture statistics,'' \em IEEE Trans. Signal Process.\em, vol.~60, no.~7, pp.~3840--3845, July~2012.
\bibitem{Banelli:13} P. Banelli, ``Bayesian estimation of a Gaussian source in Middleton's class-A impulsive noise,'' \em IEEE Signal Process. Lett.\em, vol.~20, no.~10, pp.~956--959, Oct.~2013.
\bibitem{Zhidkov:08} S. V. Zhidkov, ``Analysis and comparison of several simple impulsive noise mitigation schemes for OFDM receivers,'' \em IEEE Trans. Commun.\em, vol.~56, no.~1, pp.~5--9, Jan.~2008.
\bibitem{Golub:13} G. H. Golub and C. F. Van Loan, \em Matrix Computations\em, 4th ed. Baltimore, MD: Johns Hopkins Univ. Press, 2013.
\bibitem{Friedmann:00} J. Friedmann, H. Messer, and J. Cardoso, ``Robust parameter estimation of a deterministic signal in impulsive noise,'' \em IEEE Trans. Signal Process.\em, vol.~48, no.~4, pp.~935--942, Apr.~2000.
\bibitem{Ndo:10} G. Ndo, P. Siohan, and M. H. Hamon, ``Adaptive noise mitigation in impulsive environment: application to power-line communications,'' \em IEEE Trans. Power Del.\em, vol.~25, no.~2, pp.~647--656, Apr.~2010.
\bibitem{Yih:12} C. H. Yih, ``Iterative interference cancellation for OFDM signals with blanking nonlinearity in impulsive noise channels,'' \em IEEE Signal Process. Lett.\em, vol.~19, no.~3, pp.~147--150, Mar.~2012.
\bibitem{Tseng:12} D. F. Tseng, Y. S. Han, W. H. Mow, and L. C. Chang, ``Robust clipping for OFDM transmissions over memoryless impulsive noise channels,'' \em IEEE Commun. Lett.\em, vol.~16, no.~7, pp.~1110--1113, July~2012.
\bibitem{Alsusa:13} E. Alsusa and K. M. Rabie, ``Dynamic peak-based threshold estimation method for mitigating impulsive noise in power-line communication systems,'' \em IEEE Trans. Power Del.\em, vol.~28, no.~4, pp.~2201--2208, Oct.~2013.
\bibitem{Juwono:14} F. H. Juwono, Q. Guo, D. Huang, and K. P. Wong, ``Deep clipping for impulsive noise mitigation in OFDM-based power-line communications,'' \em IEEE Trans. Power Del.\em, vol.~29, no.~3, pp.~1335--1343, June~2014.
\bibitem{Juwono:15} F. H. Juwono, Q. Guo, Y. Chen, L. Xu, D. Huang, and K. P. Wong, ``Linear combining of nonlinear preprocessors for OFDM-based power-line communications,'' \em IEEE Trans. Smart Grid\em, vol.~7, no.~1, pp.~253--260, Jan.~2016.
\bibitem{Chuah:00} T. C. Chuah, B. S. Sharif, and O. R. Hinton, ``Nonlinear decorrelator for multiuser detection in non-Gaussian impulsive environments,'' \em IET Electron. Lett.\em, vol.~36, no.~10, pp.~920--922, 11~May~2000.
\bibitem{Delic:02} H. Deli{\c c} and A. Hocanin, ``Robust detection in DS-CDMA,'' \em IEEE Trans. Veh. Technol.\em, vol.~51, no.~1, pp.~155--170, Jan.~2002.
\bibitem{Guney:06} N. G{\"u}ney, H. Deli{\c c} and M. Koca, ``Robust detection of ultra-wideband signals in non-Gaussian noise,'' \em IEEE Trans. Microw. Theory Tech.\em, vol.~54, no.~4, pp.~1724--1730, June~2006.
\bibitem{Beaulieu:08} N. C. Beaulieu and B. Hu, ``Soft-limiting receiver structures for time-hopping UWB in multiple-access interference,'' \em IEEE Trans. Veh. Technol.\em, vol.~57, no.~2, pp.~810--818, Mar.~2008.
\bibitem{Beaulieu:10} N. C. Beaulieu and S. Niranjayan, ``UWB receiver designs based on a Gaussian-Laplacian noise-plus-MAI model,'' \em IEEE Trans. Commun.\em, vol.~58, no.~3, pp.~997--1006, Mar.~2010.
\bibitem{Rabbani:09} H. Rabbani, R. Nezafat, and S. Gazor, ``Wavelet-domain medical image denoising using bivariate Laplacian mixture model,'' \em IEEE Trans. Biomed. Eng.\em, vol.~56, no.~12, pp. 2826--2837, Dec.~2009.
\bibitem{Rabbani:10} H. Rabbani and S. Gazor, ``Image denoising employing local mixture models in sparse domains,'' \em IET Image Process.\em, vol.~4, no.~5, pp. 413--428, 2010.
\bibitem{Gersho:77} A. Gersho, ``Quantization,'' \em IEEE Commun. Mag.\em, vol.~15, no.~5, pp.~16--29, Sept.~1977.
\end{thebibliography}
\end{document}